# *Not sure?* Handling hesitancy of COVID-19 vaccines

N.F. Johnson[1,*], N. Velásquez[2], R. Leahy[2,3], N. Johnson Restrepo[2,3], O. Jha[1], Y. Lupu[4]
[1]Physics Department, George Washington University, Washington D.C. 20052
[2]Institute for Data, Democracy and Politics, George Washington University, Washington D.C. 20052
[3]ClustrX LLC, Washington D.C.
[4]Department of Political Science, George Washington University, Washington D.C. 20052
* neiljohnson@gwu.edu

**From the moment the first COVID-19 vaccines are rolled out, there will need to be a large fraction of the global population ready in line. It is therefore crucial to start managing the growing global hesitancy to any such COVID-19 vaccine. The current approach of trying to convince the "no"s cannot work quickly enough, nor can the current policy of trying to find, remove and/or rebut all the individual pieces of COVID and vaccine misinformation. Instead, we show how this can be done in a simpler way by moving away from chasing misinformation content and focusing instead on managing the "yes--no--not-sure" hesitancy ecosystem.**

Only 42 percent of Americans in the September YouGov poll[1] said "yes" to receiving a future COVID vaccine, down from May across all political sides. That means that even in a best-case scenario where a future high performing vaccine is 80% effective in an individual, it would only impact 42x80=34% of the population which is way below predicted thresholds for herd immunity. Worse, current narratives within online social media communities and media interviews[2] suggest that many of these "yes" respondents would say "no" to being first in line.

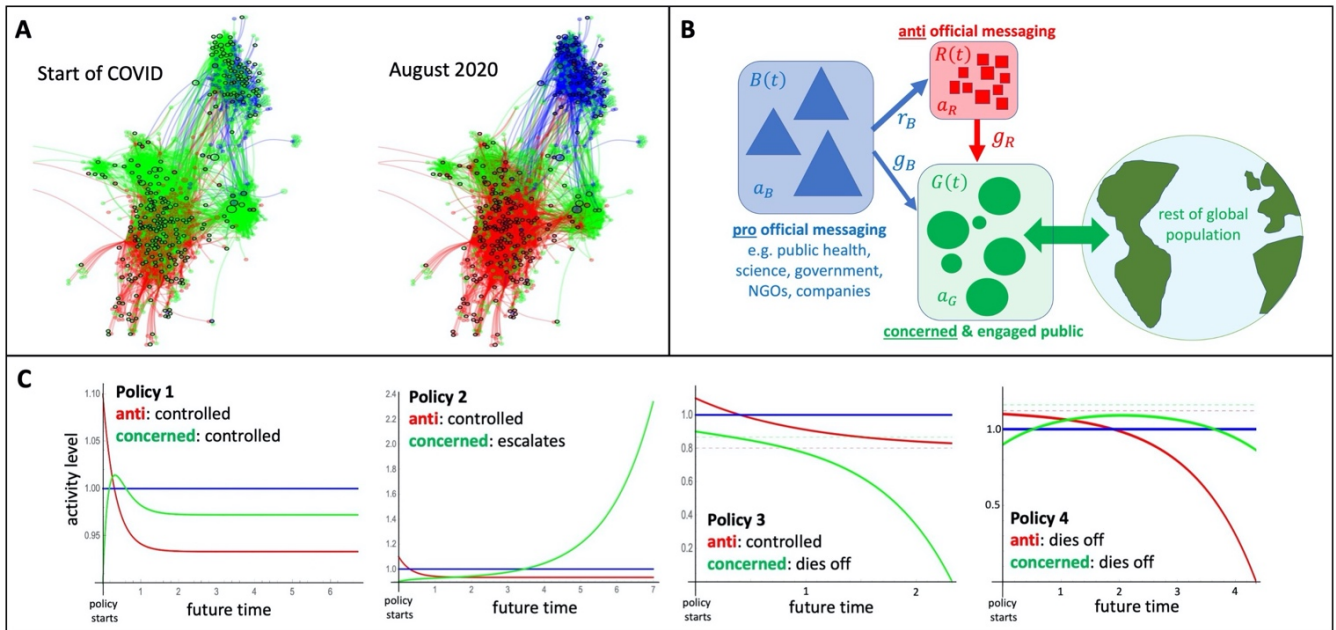

**Fig. 1: <u>Hesitancy ecosystem</u>. A: Evolution of pro (Blue), anti (Red) and "not sure" (Green) vaccine views within communities (nodes) on Facebook during COVID. Methodology and layout same as Ref. 6. Each node is a Facebook Page with 10-1,000,000+ members. Black circles show nodes that changed color, i.e. changed their vaccine view. There are currently many additional nodes which are Green but for clarity are not shown. B: Our empirically grounded ecosystem model. Each subpopulation comprises nodes of a given color from A. C: The model's four predicted futures depend on the policy choice being made, i.e. they depend on the messaging activity coupling values $(g_B, g_R, r_B)$. For illustration, the couplings here are $(1.0, 4.0, 1.0), (-2.3, 0.5, 1.0), (-2.0, -0.5, -1.0), (-0.5, 0.5, -2.5)$. Each type of outcome is an exact solution of the model in B (see Supporting Online Material SOM) and is robust in the parameter space.**

Such a high level of hesitancy risks everyone's health[3,4,5] and hopes of a 'return to normal'. It risks the diversity of volunteers for Phase III trials, and the billions of dollars allocated to vaccine development. It risks the likelihood of people accepting future COVID booster shots. And it undermines trust in existing vaccines for other diseases, as well as more general public health advice.

Trying to change the minds of hardcore "no"s[1] will be too slow and too hard. Not only has anti-vaccination sentiment been around since vaccines were first created, our study of the social media ecosystem shows that anti-vaccination communities actually grew[6] in size and resilience during 2019 -- despite the fact that measles outbreaks were proliferating throughout 2019 and despite the fact that the measles vaccine already exists and has a strong safety record. Worse, their opposition to a future COVID vaccine has now gone into overdrive (Fig. 1A, red nodes). They won't be changing their minds any time soon.

A better approach is to focus on the "not sures": and we don't just mean the 31% who specifically responded "not sure" in the latest poll[1], but also the large number of "yes" respondents who say they will delay receiving the vaccine until others have had it. Our analysis of the online Facebook ecosystem -- of which the poll appears crudely representative -- reveals well in excess of 100 million such "concerned" individuals. Each is a member of a community comprising 10-1,000,000 like-minded fans of a particular topic that is typically unrelated to vaccines, e.g. pet lovers, parent school groups, yoga fans, foodies or alternative health followers (green nodes, Fig. 1A). Since members of the same community tend to trust each other on this one topic or lifestyle choice (e.g. pet care, best choice of kindergarten, wine or organic blueberries), they also tend to listen when their community starts talking or posting about COVID and vaccines. And their growing collective concern has led their community to form links with anti-vaccination communities (red nodes, Fig. 1A).

But how to reduce the hesitancy of these "concerned" communities?

**Misinformation minefield**
Given the vicious circle of hesitancy and misinformation, one might try ramping up the current content-focused policy approach of labeling, removing or debunking specific stories. But this is not practical given the escalating number and nature of such stories and the need to act now before vaccine rollout. It may also backfire. Take the often cited story that Bill Gates is planning to put semiconductor markers into COVID vaccines, so that medical records can be scanned from our families' foreheads at school or work like a can of supermarket soup. Both elements are indeed true: semiconductor quantum dots[7] can be excellent biomarkers, and The Gates Foundation has been involved in funding related research as well as COVID vaccine development. So while there is zero likelihood this story will play out, it is in principle possible scientifically. So classifying it as 'misinformation' and removing it could spark claims of stifling free speech, while saying it is 'wrong science' amplifies the debate of what is 'right science'. Both fuel the misinformation fire and both are now rampant within the "concerned" communities (green nodes).

Worse, Facebook itself cannot find all misinformation within its own platform. Leaving pieces untouched can wrongly suggest to users that they are true. Moreover, misinformation also flows freely within and between other platforms such as 4Chan which are beyond Facebook's control[8]. So to do all this well, public health agencies and vaccine manufacturers would have to become more expert in social media than Facebook -- which is again impossible.

Then come the content flavors. The entire establishment health enterprise ("Blue", Fig. 1B) can do little more than put out statements that are scientifically correct, which means almost by definition that they are quite standard and plain: vanilla. By contrast, the anti-establishment health subpopulation ("Red", Fig. 1B) collectively offers the concerned-and-engaged subpopulation ("Green", Fig. 1B) all sorts of tempting flavors of narrative. These range from the lack of any long-term safety record for a COVID vaccine, which is of course technically true; to claims that the human immune system offers a superior form of resistance, which is also hard to disprove given the lack of understanding COVID biology; to claims of hidden agendas of governments and big pharma, which again is hard to disprove given the highly political nature of the COVID vaccine race and the billions being invested to secure future batches; to the fact that science is still struggling to give precise answers to seemingly straightforward (but actually highly complex) questions such as best-choices for school opening hours, numbers in a class, and mask design.

Blue cannot hope to win such a content-chasing war quickly enough for vaccine rollout. It would get bogged down in virtual whack-a-mole across the ever-expanding multiverse of interconnected social media platforms, with an ever-expanding set of stories to tackle as different vaccine candidates from different countries come online and different political decisions loom.

**More when and less what**
Here we suggest a more immediate and less resource-intensive approach that leverages what pro-establishment messaging (Blue) already has at its disposal, i.e. when to engage in messaging and at what level in terms of volume. Red and Green feed off of Blue's activity, and Green also feeds off of Red (Fig. 1B). Figure 1C shows how powerful the approach of managing the ecosystem could be using a simple, undergraduate level model (Fig. 1B) that combines Newman's gossip model[9] and Strogatz's relationship dynamics model[10], and which is backed up by empirical findings and theory from studies of online opinion formation[11] and conflicts[12] as well as Ref. 6. There are very few parameters (Fig. 1B) and each is physically meaningful. Most importantly, this simple model reproduces the main features of the evolution of the Reds, Blues and Greens from the start of COVID until now (see SOM).

The advantages of this type of ecosystem analysis for policymakers are that the predictions are available as precise, plug-and-play formulae (see SOM) that can be evaluated by hand or using a simple calculator app on a phone. No computer simulations or coding required. Interestingly, it predicts 4 distinct classes of outcomes and hence 4 classes of policy (Fig. 1C), each of which could be implemented immediately as discussed below.

**Policy predictions**
As with any other issue such as climate change, framing policy discussions in terms of calculated behaviors with quantifiable and testable predictions, is far more powerful than vaguer verbal arguments about what might work. Immediate action is essential: if left to carry on as is, Fig. 1A predicts Red's stronghold will not only continue growing, it will draw in and likely tip an increasing fraction of Greens which will seriously undermine all future COVID vaccine rollouts and renewals.

Policies 1-4 show the trade-offs for Blue: on the one hand, Blue must get communities such as pet lovers and yoga fans away from concern about COVID and vaccines, and back to back to their real interests, i.e. reduce the green curve in Fig. 1C. At the same time, it must keep the red curve under control (Red). And it must do all this using the blunt instrument of its own messaging activity and without feeding the infodemic frenzy, i.e. Blue's average output must remain steady (blue curve).

**Policy 1** shows what happens if Blue mirrors Red and Green's messaging activity, in the scenario that Red and Green are also doing the same to each other, i.e. the couplings in Fig. 1B are all positive ($g_B, g_R, r_B > 0$). This is like becoming louder when the other is loud, and quieter when the other is quiet. The "concerned"s green curve initially increases and hence gets worse, but then settles to a stable value. Red activity drops to a lower steady value. While not a dramatic improvement in overall hesitancy, it is in principle possible to choose the positive coupling values such that total support is above the estimated herd immunity threshold. Policymakers would however have to warn the public that things will initially feel slightly worse before improving.

**Policy 2** differs from Policy 1 in that it shows what happens if Blue has high messaging activity levels when Green is low and vice versa ($g_B < 0$). This is like becoming louder when the other is quiet, and quieter when the other is loud. Though Red activity drops which is desirable, Green escalates dramatically which is undesirable.

**Policy 3** is the opposite of Policy 1 in that Blue, Red and Green all have negative feedback ($g_B, g_R, r_B < 0$). So they become louder when the others are quiet and vice versa. The outcome is good all round: not only does Red activity drop to a steady state, Green drops dramatically and keeps decreasing over time.

**Policy 4** differs from Policy 3 in that Red and Green now have positive feedback ($g_R > 0$), so Red and Green become louder when the other is loud, and quieter when the other is quiet. This policy has the advantage over Policy 3 in that both Red and Green eventually both keep decreasing over time. However Green initially gets worse, before then showing a turning point which can be calculated and hence predicted exactly.

Which of these 4 policies is most suitable will depend on the current value of the couplings in Fig. 1B. But the key is that they can each be analyzed and compared ahead of implementation on a case-by-case basis depending on Blue's level of control of the various parameters in Fig. 1B, using the plug-and-play formulae in the SOM. Hence the system can be nudged toward the estimates required for herd immunity.

**Beyond vaccines**
These policies also apply to other situations where there is competition between establishment messaging (Blue), anti-establishment messaging (Red) and a background population (Green) whose 'hearts and minds' can tip the balance[13,14]. For example, it could help with the contentious climate-change narratives that are circulating concerning the September 2020 wildfires in California -- and it could kick-start the needed public engagement before quantum information technologies are unleashed[15]. Also the analysis doesn't just apply to hesitancy online. Working with epidemiologists, more detail can be added to Fig. 1B by incorporating details of how communities are interconnected within each subpopulation and how the messaging spreads, hence yielding a fuller theory of infodemic spreading within a heterogeneous population[16,17]. Moreover, the role of specific content could be included using machine learning[18], with different types of misinformation having different coupling values in Fig. 1B, while cleverer use of human psychology could enhance the model's realism[19,20,21].

**Supporting Online Material (SOM)**

1. Overview of the equations used for the example outputs in Fig. 1C

$\dot{R} = a_R(R_0 - R) + r_B(B - R)$
$\dot{B} = a_B(B_0 - B)$ so fixed point $B^* = B_0$ always
$\dot{G} = a_G(G_0 - G) + g_R(R - G) + g_B(B - G)$

Dynamics $(R(t), B(t), G(t))$ which is amount of COVID-19 discussion in Red, Blue, Green at time *t*
Initial conditions $(R(0), B(0), G(0))$
Fixed point $(R^*, B^*, G^*)$

$\epsilon_R(t) = R(t) - R^*$, $\epsilon_B(t) = B(t) - B^*$, $\epsilon_G(t) = G(t) - G^*$

To predict policy impacts (Fig. 1C) we then simply run the model forward using realistic parameters for the current instant in time. For simplicity here, let's suppose that Blue will continue to put out scientific messaging and advice as the COVID vaccine research develops. This output does not depend on the gossip going on within Red and Green, hence the direction of the arrow and coupling in the model. Green absorbs this (see SI) and to some extent so does Red, but Red also has its own activity toward Blue and Green. So the couplings $g_R$ and $r_B$ (Fig. 1B) are largely in Red's control, not in Blue's. Hence Fig. 1C focuses on Blue only being able to control its activity level with respect to Green, $g_B$, and then only in some limited way. A positive coupling between two populations means they will tend to synchronize their activity, while negative means they will tend to become out of sync.

2. Proof that the model reproduces the features in the empirical data during 2020, despite have very few parameters.

Shown here is the actual data (left panels) and model predictions (right panels) for the number of communities (i.e. clusters, each of which is a node in Fig. 1A) for (top) the number of clusters that are subscribers to other clusters that are broadcasting COVID narratives and hence 'listening to COVID narratives', and (bottom) the number of clusters that are broadcasting COVID narratives to other clusters and hence 'talking COVID narratives'.

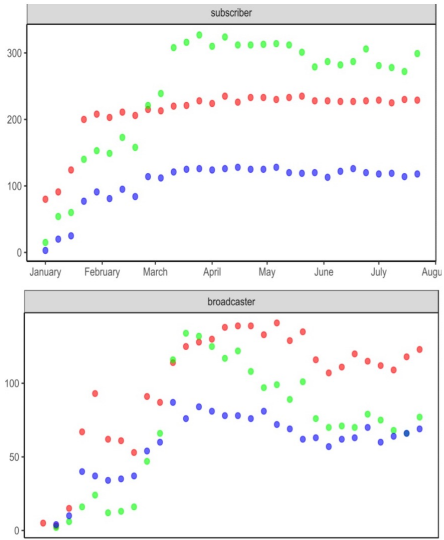
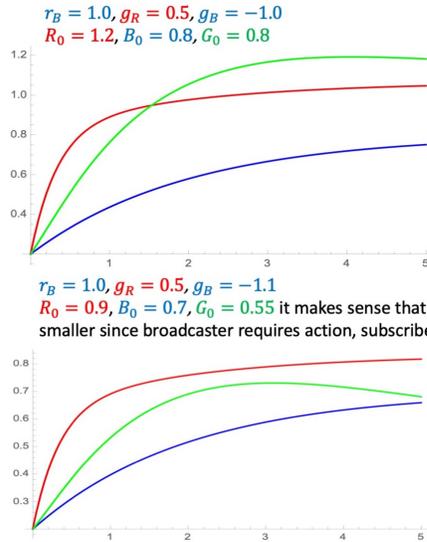

## 3. More details about the predicted policy result in Fig. 1C

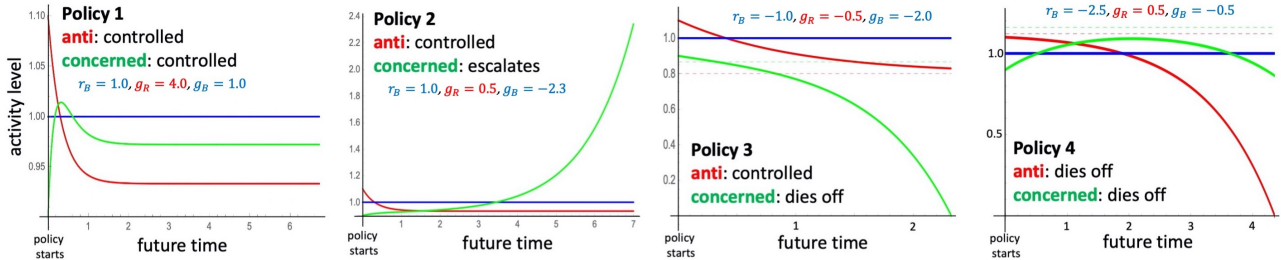

Here we suppose that we are starting a new policy from a new moment in time in the near future when vaccines getting closer, and so after Blue has strengthened and Red and Green have adjusted their capacity e.g. $R_0 = 0.9$, $B_0 = 1.0$, $G_0 = 1.1$

We start from this point in time with initial conditions $(R(0), B(0), G(0)) = (1.1, 1.0, 0.9)$ because Red starts in the lead (even though has smaller long-term capacity $R_0$) while Green is still relatively quiet but has a larger capacity since it can in principle include all the concerned public

The 4 types of trajectories start near $(R^*, B^*, G^*)$ with $B(t) = B^*$ for all time. As always, $a_R = 2.0$, $a_B = 0.5$, $a_G = 1.0$ directly from the empirical data of 2019

## 4. The exact formulae that produce the predicted policy results in Fig. 1C

Consider Red ($R$), Blue ($B$) and Green ($G$) subpopulations. Internally, each subpopulation is a partially connected set of nodes where each node is an online community such as a Facebook Page. This internal structure of each subpopulation does not concern us here since we are analyzing a model of interacting subpopulations where the unit of analysis is the subpopulation. The coupling of $G$ to the rest of the world is some outside term which does not concern us. The dynamics coupling these subpopulations is described by the following time-dependent equation $\dot{x}_i = f(x_i) + \sum_j A_{ij} C_{ij}(x_i, x_j)$ where $A_{ij}$ has components $a_{ij}$ to represent the net links from $i$ to $j$ and $C_{ij}(x_i, x_j)$ is the dynamical coupling. Our analysis of the empirical data from 2019 in our earlier paper, suggest the following forms of equations. We stress that although these appear as very specific choices, the results are more general since it is not the precise form that is important but rather the topological features of the functions. We will take the linear expansion of the coupling terms to first order, i.e. $C_{ij}(x_i, x_j) \sim g_{ij}(x_i - x_j) + \ldots$ where $g_{ij}$ is the coupling constant:

$$\begin{aligned}
\frac{dR}{dt} &= a_R(R_0 - R) + r_B(B - R) \\
\frac{dB}{dt} &= a_B(B_0 - B) \\
\frac{dG}{dt} &= a_G(G_0 - G) + g_R(R - G) + g_B(B - G)
\end{aligned} \quad (1)$$

It makes sense that the coupling term, e.g. from Red into Green, will depend in some way on the difference between Red and Green activity levels $R(t)$ and $G(t)$. Positive couplings $r_B$, $g_R$ and $g_B$ imply positive feedback, e.g. $r_B > 0$ means that excess activity in $B$ compared to $R$ will increase the activity in $R$ by increasing the value of rate of change of $R(t)$ and so on. Likewise, $r_B < 0$ implies negative feedback so that excess activity in $B$ compared to $R$ will decrease the activity in $R$ by decreasing the value of rate of change of $R(t)$. The stable fixed point here is $(R^*, B^*, G^*)$. The expressions are:

$$\begin{aligned}
R^* &= \frac{a_R R_0 + r_B B_0}{(a_R + r_B)} \\
B^* &= B_0 \\
G^* &= \frac{a_G G_0 + g_B B_0}{(a_G + g_R + g_B)} + \frac{g_R(a_R R_0 + r_B B_0)}{(a_R + r_B)(a_G + g_R + g_B)} .
\end{aligned} \quad (2)$$

Taking $\epsilon = (\epsilon_R, \epsilon_B, \epsilon_G) = (R - R^*, B - B^*, G - G^*)$ as the deviation from the stable point $(R^*, B^*, G^*)$, we obtain:

$$\begin{aligned}
\frac{d\epsilon_R}{dt} &= -a_R \epsilon_R + r_B(\epsilon_B - \epsilon_R) \\
\frac{d\epsilon_B}{dt} &= -a_B \epsilon_B \\
\frac{d\epsilon_G}{dt} &= -a_G \epsilon_G + g_R(\epsilon_R - \epsilon_G) + g_B(\epsilon_B - \epsilon_G)
\end{aligned} \quad (3)$$

The eigenvalues are:

$$\begin{aligned}
\lambda_1 &= -a_B \\
\lambda_2 &= -(a_R + r_B) \\
\lambda_3 &= -(a_G + g_R + g_B)
\end{aligned} \quad (4)$$

Consider the case going forward when $B$ remains at its fixed point at all times ($\epsilon_B = 0$). This means that on average $B$'s messaging activity level remains essentially constant which reflects the fact that scientific information flows in at a steady state, and so the flow from Blue does not depend on what $R$ or $G$ are saying or thinking. The dynamics essentially reduces to a two dimensional case:

$$\frac{d\epsilon_R}{dt} = \lambda_2 \epsilon_R$$
$$\frac{d\epsilon_G}{dt} = \lambda_3 \epsilon_G + g_R \epsilon_R$$
and so in matrix form, $\quad \frac{d\boldsymbol{\epsilon}}{dt} = \mathcal{M}\boldsymbol{\epsilon} \quad$ (5)

The eigenvalues and the corresponding eigenvectors of $\mathcal{M}$ are $\lambda_2$, $\lambda_3$ and $\begin{pmatrix} \lambda_2 - \lambda_3 \\ g_R \end{pmatrix}$, $\begin{pmatrix} 0 \\ 1 \end{pmatrix}$ respectively. Writing $\boldsymbol{\epsilon}$ as a linear combination of the eigenvectors of $\mathcal{M}$, we have

$$\boldsymbol{\epsilon}(t) = C_1 e^{\lambda_2 t} \begin{pmatrix} \lambda_2 - \lambda_3 \\ g_R \end{pmatrix} + C_2 e^{\lambda_3 t} \begin{pmatrix} 0 \\ 1 \end{pmatrix} \quad (6)$$

The corresponding solutions for the deviation from the fixed points $R*$ and $G*$ are:

$$\epsilon_R(t) = \epsilon_R(0) e^{\lambda_2 t}$$
$$\epsilon_G(t) = \frac{g_R \epsilon_R(0)}{\lambda_2 - \lambda_3} e^{\lambda_2 t} + \left[\epsilon_G(0) - \frac{g_R \epsilon_R(0)}{\lambda_2 - \lambda_3}\right] e^{\lambda_3 t} \quad (7)$$

Hence

$$R(t) = \frac{a_R R_0 + r_B B_0}{(a_R + r_B)} + \epsilon_R(0) e^{\lambda_2 t}$$
$$G(t) = \frac{a_G G_0 + g_B B_0}{(a_G + g_R + g_B)} + \frac{g_R(a_R R_0 + r_B B_0)}{(a_R + r_B)(a_G + g_R + g_B)} + \frac{g_R \epsilon_R(0)}{\lambda_2 - \lambda_3} e^{\lambda_2 t} + \left[\epsilon_G(0) - \frac{g_R \epsilon_R(0)}{\lambda_2 - \lambda_3}\right] e^{\lambda_3 t}$$
(8)